\documentclass{article}
\usepackage{ijcai22}

\usepackage{times}
\usepackage{soul}
\usepackage{url}
\usepackage[hidelinks]{hyperref}
\usepackage[utf8]{inputenc}
\usepackage[small]{caption}
\usepackage{graphicx}
\usepackage{amsmath}
\usepackage{amsthm}
\usepackage{booktabs}
\usepackage{algorithm}
\usepackage{algorithmic}
\usepackage{natbib}

\title{Tell Me Something That Will Help Me Trust You:\\
A Survey of Trust Calibration in Human-Agent Interaction 
}

\author{
George J. Cancro$^{1,2}$\and
Shimei Pan$^1$\And
James Foulds$^1$
\affiliations
$^1$University of Maryland, Baltimore County, Information Systems Department\\
$^2$Johns Hopkins University Applied Physics Laboratory, Laurel, MD \\
\emails
\{gcancro1, shimei, jfoulds\}@umbc.edu
}

\begin{document}

\maketitle

\begin{abstract}
When a human receives a prediction or recommended course of action from an intelligent agent, what additional information, beyond the prediction or recommendation itself, does the human require from the agent to decide whether to trust or reject the prediction or recommendation? In this paper we survey literature in the area of trust between a single human supervisor and a single agent subordinate to determine the nature and extent of this additional information and to characterize it into a taxonomy that can be leveraged by future researchers and intelligent agent practitioners. By examining this question from a human-centered, information-focused point of view, we can begin to compare and contrast different implementations and also provide insight and directions for future work.
\end{abstract}

\section{Introduction}

Imagine you are a Hearing Examiner conducting a parole hearing and it is your job to decide whether an inmate should be released from prison. To increase consistency and improve decision-making, you have been provided a partner; an intelligent agent who, upon reviewing the inmate’s record and other attributes, provides a recommendation to either approve or refuse parole. In one specific case, you recommend parole, but the intelligent agent recommends parole be refused. Now what do you do?

The situation you are in is difficult because even though the intelligent agent was presented as your partner, you have a legal and ethical responsibility to get this decision right. Therefore, you are acting as the agent’s supervisor; faced with a decision to either override or accept your subordinate’s (the agent’s) recommendation. 

This decision is called a ``\emph{trust calibration}'' decision. We define trust calibration as a process by which a human supervisor determines whether an individual agent’s prediction or recommended course of action should be trusted on a case-by-case basis. The original work on trust calibration was performed by \citet{Muir1987Trust} who stated: the human supervisor ``\emph{must learn to calibrate his trust, that is to set his trust to a level corresponding to [an agent’s] trustworthiness, and then use the [agent] accordingly.}'' 

Unfortunately, not all human supervisors ‘use the agent accordingly’. Some humans just side with the intelligent agent (i.e., ``automation bias'' or ``missed intervention'') \citep{Lyell2017Automation,Lee2018Hulu}. Others immediately override the recommendation and go with their own decision (i.e., ``algorithm aversion'' or ``bad intervention'') \citep{Dietvorst2015AlgorithmAP,Awad2018BlamingHI}. 
In a benign situation (e.g., movie recommendation), being automation biased or algorithm adverse may have little consequence.  
This is not the case in high-stakes situations (e.g., military, security, health-care, criminal justice), where inability to calibrate trust correctly can result in severe effects to human life.  
For example, for algorithm aversion, if judges had agreed with an intelligent agent in 'failure to appear' decisions, pre-trial detention rates would have been reduced by 41.9\% with no increase in crime rate \citep{Lowens2020AlgAver}. 
In another example, for algorithm bias, if operators had not trusted the Patriot missile targeting system course of action, a Royal Air Force jet would not have been shot down by friendly fire \citep{Bode2021AlreadyWrong}.

The ideal partnership lies between these two extremes by achieving the goal of complementary performance; where the human supervisor agrees with the agent when the agent's performance is likely to exceed the supervisor, and the supervisor rejects the agent when the agent is likely to under-perform.

To determine when the agent is likely to exceed or under-perform the human supervisor, you will need more information than simply the main output of the agent. We call this additional information ``\emph{trust calibration meta-information}.'' We define trust calibration meta-information as information flowing from the agent subordinate to the human supervisor that provides context to the main output of the agent; thereby facilitating a better understanding of whether the main output should be accepted.

The concept of trust calibration appears in many different research fields: automation, decision-making, recommender systems, human-robot interaction, and explainable AI (XAI). In some fields, such as automation \citep{Lee2004TrustIA}, decision-making \citep{Wang2021AreEH} and recommender systems \citep{Kunkel2019LetME}, trust calibration has been apart of the literature for some time. In human-robot interaction, trust includes multiple factors related to performance, appearance, and interaction behavior \citep{Khavas2021ARO}, but trust calibration decisions still exist to determine whether the human supervisor should intervene in a robot's course of action \citep{Wang2016TrustCW}. In XAI, much focus is given to understandability, comprehensibility, interpretability, and transparency \citep{Arrieta2020ExplainableAI} but a ``central task for XAI is calibrating trust'' \citep{Sperrle2020ShouldWT}. To ensure coverage across this wide set of research, we structured our survey as a trace through these various fields.

\subsection{Test Set Accuracy vs. Meta-Information}

When confronted with the decision to accept or intervene in an agent main output, one might consider the test set accuracy of the intelligent agent as a useful guide. Test set accuracy (i.e., the performance of the agent on a set of data not used during agent training) can be considered trust calibration meta-information, but in most cases is not sufficient for this decision. For example, knowing an agent achieved 99\% accuracy during testing, does not help determine if the current case is the statistical 1 out of 100 where the agent is incorrect. And moreover, test set accuracy does not inform the user if the current case is even in the same distribution as the cases used during the agent’s training or testing. Therefore, we need to consider what additional types of meta-information can be leveraged.

\subsection{Why is a Taxonomy Needed?}

As this survey will show, multiple works have considered trust calibration and have identified, indirectly, the meta-information associated with their proposed methods. Unfortunately, it is difficult to cross-compare these papers since terminology used is overlapping in some senses and conflicting in others. 
For example, 
\citet{Zhang2020EffectOC} states ``unlike confidence, explanation did not seem to affect participants' trust.''
In another paper, 
\citet{McGuirl2006SupportingTC} state ``confidence information can improve trust calibration.''
Upon closer examination, even though these authors appear to agree, these authors have used the same term, 
``confidence'' with slightly different meanings.
Zhang uses ``confidence'' to refer to historical performance values, 
while
McGuirl uses ``confidence'' to refer to an uncertainty of how close the current case matches the training data set. 

Before making judgements on what types or modalities of trust calibration meta-information best drives human-agent partnerships toward the goal of complementary performance, we believe the community needs a taxonomy to specifically characterize the meta-information leveraged in the different literature on trust calibration.

Therefore, the goal of this paper is to survey trust calibration from a meta-information point of view and create a taxonomy of trust calibration meta-information; thereby providing a more rigorous basis for describing and contrasting different approaches to trust calibration. This taxonomy should improve the discussion in the field of trust calibration, lead to more nuanced and effective comparison studies across different trust calibration approaches, and potentially open new areas of trust calibration research.

In summary, the novel contributions of our work include:
\begin{itemize}
\item A taxonomy for describing trust calibration meta-information, in terms of content, modality, interactivity, and timing.
\item An overview of trust calibration literature from a human-computer interaction point of view
\item Identification of future research opportunities in the area of trust calibration meta-information.
\end{itemize}

The remainder of the paper is structured as follows: Section 2 reviews the work done in performing the survey.  Section 3 details the taxonomy and demonstrates how surveyed literature is characterized within the taxonomy. In Section 4 we provide a discussion of initial findings generated by leveraging our taxonomy. Finally, we present limitations and conclusions in Sections 5 and 6 respectively.

\section{Survey Methodology}
Our survey on trust calibration leveraged a bootstrapping method. Starting with a set of initial seed papers retrieved using  keywords 'trust,' 'trust calibration,' and 'calibrate trust' on Google Scholar, we surveyed papers across the automation, XAI,  decision-aids, and robotics fields and then expanded our survey by adding relevant papers through paper references. To maintain scope within our survey, we focused on literature that examined trust on an agent's instance-level decisions. We also limited our survey to papers where trust calibration is performed within an asymmetric partnership between a single human and a single agent where the human is the supervisor, and the intelligent agent is the subordinate (or assistant). Finally, instead of focusing on methods of trust calibration, we focused on literature that directly or indirectly discussed the meta-information supporting the trust calibration. This resulted in a more human-computer interaction (HCI) focused perspective that we believe will facilitate our ability to group research similarities across diverse fields and diverse agent implementations.

\section{Meta-Information Taxonomy}
After performing the survey, we attempted to classify the meta-information into major categories. In this paper, we characterize trust calibration meta-information along four dimensions: \textit{content}, \textit{modality}, \textit{interactivity}, and \textit{timing}. Content refers to the different types of information content that can be present in the meta-information. Modality refers to the different ways in which the meta-information content can be presented. Interactivity refers to whether the human supervisor can interact with the meta-information. Timing refers to the different times in which the flow of meta-information can be initiated.

Within each major category, we used results from the survey to build up sub-categories. The names of the sub-categories were generated by selecting terms that were common in multiple papers or that we felt best represented the entire type.

The combination of the major categories (content, modality, interactivity, timing) with the various sub-categories forms our proposed taxonomy for trust calibration meta-information (Table \ref{tab:taxonomy}).

Using this taxonomy, any trust calibration approach can be characterized by selecting a sub-category from each of the four major categories. For example, if an agent provides a numeric confidence as trust calibration meta-information with its main output, the associated meta-information would be classified in the above taxonomy as: \textit{Confidence/Risk Content}, \textit{Values Modality}, \textit{No Interactivity}, and \textit{Concurrent Timing}.

The following four subsections describe each of the major categories: \emph{content}, \emph{modality}, \emph{interactivity}, and \emph{timing}.

\begin{table}
\centering
\begin{tabular}{rl}
\hline
Major Category  & Sub-Category  \\
\hline
\textbf{Content}     & Case-Based     \\
            & Sensitivity      \\
            & Influence     \\
            & Confidence/Risk     \\
            & Historic Performance     \\
            & Uncertainty     \\
            & Logic Rationale     \\ [4pt]
\textbf{Modality}	& Narriative     \\
            & Values     \\
            & Graphical     \\ [4pt]
\textbf{Interactivity}	& Yes/No     \\ [4pt]
\textbf{Timing}	    & Pro-active     \\
            & Concurrent     \\
            & Post-hoc     \\
\hline
\end{tabular}
\captionsetup{justification=centering}
\caption{Trust calibration meta-information taxonomy structure of major categories and sub-categories}
\label{tab:taxonomy}
\end{table}

\subsection{Meta-Information Content}
Although there is a huge diversity of papers in trust calibration, when we examine the content of meta-information, we observed 7 meta-information content sub-categories, as shown in Table \ref{tab:contentSurvey}, that are common across previous works. We summarize the multiple sub-categories of meta-information content below. 

\subsubsection{Case-based Content}
\emph{Case-based content} uses examples (positive or negative) from the agent training, validation, or operations that is pertinent to the current decision or situation in order to provide the human supervisor with context of how previous decisions were made or situations were handled. By understanding the relationship between the current situation and the past examples, the human supervisor has more information by which to judge the agent’s ability.

Case-based content typically is presented by showing the prior case that is most similar to the current situation \citep{Lai2019OnHP}. If more than one case contains the same input as the current situation, the percentage of exact matches for each output class can be shown \citep{Dodge2019ExplainingMA}.  If no exact matches are found, a nearest neighbor process (e.g., a KNN approach) can be run to determine the closest match.  More than one closest match can also be provided. For example,  \citet{Lai2019OnHP,Cai2019TheEO,Wang2021AreEH} provide multiple cases: the closest match from the training data that has the same output as the current situation and the closest match that has the outputs that differ from the current situation. Case-based content can also take the form of a comparison between the current situation and a more accepted case \citep{Gregor1999ExplanationsFI,Bussone2015TheRO,Ehrlich2011TakingAF}. For example, \citet{Bussone2015TheRO} suggested that comparison of current symptoms to accepted symptoms from a standard medical diagnosis can provide the context for agent's diagnosis. The comparison to the accepted case can be side-by-side or through the calculation of a similarity metric \citep{Ehrlich2011TakingAF}. 

\subsubsection{Sensitivity Content}
\emph{Sensitivity content} presents the human supervisor with how much one or more inputs of the task would have to differ to change the output. Sensitivity content attempts to refine the human supervisor’s mental model of the agent by providing insight into the agent’s decision boundaries. Decision boundaries represent a more compact representation of the details of the agent’s decision-making process and enables the human to compare the agent’s decision boundaries to their own.

Sensitivity content is typically presented as what is the input change necessary to change the output \citep{Dodge2019ExplainingMA,Wang2021AreEH} (e.g., ``If the defendant's age had been 29 instead of 26, she would have predicted as NOT likely to re-offend'') \citep{Wang2021AreEH}. 

\subsubsection{Influence Content}
\emph{Influence content} presents the human supervisor with how much inputs of the task influence the output. Influence content attempts to assist the human supervisor by provide insight into the agent’s decision-making process. The weights each input/feature are a pseudo-representation of the decision-making process that enable the human supervisor to make contrasting examinations of the weights by comparing them to the supervisor’s own mental model of what is most important about the task.

Influence content can be presented as: the contribution or importance of an attribute in determining the outcome \citep{Lai2019OnHP,Dodge2019ExplainingMA,Cheng2019ExplainingDA,Gregor1999ExplanationsFI,Wang2021AreEH} (e.g., showing the level of importance of each word in a product review to determine if review is genuine \citep{Lai2019OnHP}), or as an indication of which inputs are impacting the outcome or course of action \citep{Wintersberger2021EvaluatingFR,Ehsan2021ExpandingET,Wang2016TrustCW,Bussone2015TheRO,Svrcek2019TowardsUP} (e.g., ``these objects in the environment are being considered in an automated vehicles driving decisions'' \citep{Wintersberger2021EvaluatingFR}.)

\subsubsection{Confidence/Risk Content}
\emph{Confidence/Risk content} is the estimate, by the agent, of the likelihood of an event occurring (task success, correct answer, loss of life, etc).   Confidence/Risk content is a higher-level content than others since this content offers a summarization of the outcome of the task. While the easiest for human supervisors to relate to and make decisions based upon, it may not be the easiest to understand unless the meaning of the confidence number and/or how it was derived are clearly communicated \citep{Bussone2015TheRO}. 

Confidence is presented typically as a risk or confidence about a specific instance of an output \citep{Ashoori2019InAW,Wang2016TrustCW,Bussone2015TheRO}(e.g., ''I am 67\% confident about this assessment'' \citep{Wang2016TrustCW}).

\subsubsection{Historic Performance Content}
\emph{Historic Performance content} is success statistics on prior training, validation or operations that is pertinent to the current situation. Similar to Case-based content in terms of using prior examples, Historic Performance differs from Case-based in that Historic Performance is an aggregate measure of the past rather than providing key cases. In addition, Historic Performance content differs from Case-based in terms of purpose. Where the purpose of Case-based is to provide details on agent decision making under similar circumstances, Historic Performance focuses on whether the agent should be used in this situation or not based on aggregated statistics.

Historic Performance is typically presented as performance during validation or past operational experience (e.g., ``Agent has an accuracy of approximately 87\%'') \citep{Lai2019OnHP,Wang2016TrustCW}, but also can be presented as performance against similar inputs in the past (e.g., ``The agent is correct N times out of 10 in cases similar to this one'') \citep{Zhang2020EffectOC,Dodge2019ExplainingMA}, performance against other peers in operation (e.g., 1 out of 3 peers sold at the recommended price) \citep{Ehsan2021ExpandingET}, or performance for attribute-based groups (e.g., ``58\% of people in 18-29 age group re-offended'') \citep{Dodge2019ExplainingMA}. 

\subsubsection{Uncertainty Content}
\emph{Uncertainty content} is an estimate, by the agent, of the lack of knowledge about the outcome. Uncertainty attempts to assist the human supervisor by providing insight how well the agent understands the problems represented by the task. For example, if an agent’s uncertainty is perceived by the human supervisor as too large, the supervisor may be less likely to trust the outcome of the agent \citep{Bhatt2021UncertaintyAA}. Uncertainty can be broken down into aleatoric and epistemic uncertainty. Aleatoric uncertainty (doubt) represents uncertainty inherent in the system being modeled because of stochastic behavior. Epistemic uncertainty (ambiguity) is the uncertainty due to limited data or knowledge \citep{Tomsett2020RapidTC}. Uncertainty can be presented as the error bars on an outcome, probability distribution summary, or spaghetti plots \citep{Bhatt2021UncertaintyAA}. 

Another important area of uncertainty is the concept of out-of-distribution. Out-of-distribution refers to difference in inputs that occur when an agent is trained and validated with a closed-world assumption (all data from the same distribution) and when the agent is deployed into an open-world scenario where inputs can now be out-of-distribution from the training and validation data sets \citep{Yang2021GeneralizedOD}. A form of epistemic uncertainty, out-of-distribution detection attempts to determine how far new inputs are from the inputs that the agent was trained on \citep{Tomsett2020RapidTC}. By providing an estimate if the current data is out-of-distribution, an agent provides a powerful estimate of uncertainty. This type of uncertainty content can be presented as a percentage of how close the current input matched the training set \citep{McGuirl2006SupportingTC} and is a great counterbalance to Confidence/Risk content as it provides a better understanding of how much weight to put on agent-generated confidence (e.g., a high confidence should be viewed as useless if the agent also provides high probability that the assessment was performed on inputs out-of-distribution with training inputs).

\subsubsection{Logic Rationale Content}
\emph{Logic Rationale} content provides insight into the inner workings of agent logic that led to the outcome. The benefit of Logic Rationale content is that it assists the human supervisor to understand the “why” behind agent actions.

Logic Rationale is typically presented in the form of limited logic statements \citep{Ehsan2021TheWI,Zhu2020EffectsOP,Svrcek2019TowardsUP} (e.g., ``I am going to collect red resources because you are low on red resources'' \citep{Zhu2020EffectsOP}, or ``This was recommended because it contains features of movies you positively rated in the past'' \citep{Svrcek2019TowardsUP}). Logic Rationale can also be presented in the form of actual internal logic, if the agents method allows (e.g., decision tree, rules) \citep{Huysmans2011AnEE} or if the understandable logic can be extracted from a black box agent \citep{Guidotti2019ASO}.

\begin{table*}[t]
\centering
\begin{tabular}{p{0.125\linewidth} p{0.25\linewidth} p{0.3\linewidth} p{0.2\linewidth}}
\hline
\textbf{Sub-Category} & \textbf{Description}	& \textbf{Explanation Examples} & \textbf{Relevant Papers} \\
\hline
\textbf{Case-Based} &	
Examples (positive or negative) from the agent training, testing or operations that is pertinent to the current situation &	
\textit{``The closest match to this situation in my training set is **this** and the outcome of that training example was a success.''} & 
\scriptsize\citep{Wang2021AreEH,Dodge2019ExplainingMA,Cai2019TheEO,Lai2019OnHP,Bussone2015TheRO,Ehrlich2011TakingAF,Gregor1999ExplanationsFI} \\ [4pt]

\textbf{Sensitivity}	& 
Show how much one or more inputs of the task would have to differ to change the outcome & 
\textit{``The weight of the object would have to grow by 250 lbs. before I would not be able to carry it.''}
& \scriptsize\citep{Wang2021AreEH,Dodge2019ExplainingMA}\\ [18pt]

\textbf{Influence} &	
Show how much inputs of the task influence the outcome & 
\textit{``The inputs most driving this decision was prior convictions and age.''} & 
\scriptsize\citep{Wintersberger2021EvaluatingFR,Zhang2020EffectOC,Wang2021AreEH,Dodge2019ExplainingMA,Lai2019OnHP,Cheng2019ExplainingDA,Svrcek2019TowardsUP,Wang2016TrustCW,Bussone2015TheRO,Gregor1999ExplanationsFI} \\ [3pt]

\textbf{Confidence/ Risk}	& 
Estimate by the agent of the likelihood of a correct output or an event occurring  & 
\textit{``I estimate a 57\% risk of damage to my equipment in the process of performing this task.''} &
\scriptsize\citep{Zhu2020EffectsOP,Ashoori2019InAW,Wang2016TrustCW,Bussone2015TheRO} \\ [12pt]

\textbf{Historic \newline Performance}	 & 
Success statistics on prior situations that is pertinent to the current situation & 
\textit{``I have performed this mission in similar conditions before, and did so successfully 80\% of the time.''} & 
\scriptsize\citep{Zhang2020EffectOC,Dodge2019ExplainingMA,Ehsan2021ExpandingET,Lai2019OnHP,Wang2016TrustCW} \\ [12pt]

\textbf{Uncertainty}	& Estimate by the agent of the lack of knowledge about the outcome & 
\textit{``The object in this image has a 20\% chance of representing an object I have not been trained on.''} & 
\scriptsize\citep{Tomsett2020RapidTC,Bhatt2021UncertaintyAA,McGuirl2006SupportingTC} \\ [18pt]

\textbf{Logic \newline Rationale}	& 
Provide insights into the inner workings of agent logic that led to the outcome &	
\textit{``I am going to collect red resources because you are low on red resources.''} & 
\scriptsize\citep{Ehsan2021TheWI,Zhu2020EffectsOP,Svrcek2019TowardsUP} \\

\hline
\end{tabular}
\captionsetup{justification=centering}
\caption{Summary of meta-information content sub-categories with description and explanation examples}
\label{tab:contentSurvey}
\end{table*}

\subsection{Meta-Information Modality}

Previous work in trust calibration combined aspects of meta-information flow content and modality. We offer a different analysis where we separate content and modality. The sections below summarize the multiple sub-categories of meta-information modalities.

\subsubsection{Narrative Modality}

The \emph{Narrative modality} is a text-based presentation of meta-information flow content. This modality is typically represented as one or more sentences of text \citep{Dodge2019ExplainingMA,Lai2019OnHP,Zhang2020EffectOC,Svrcek2019TowardsUP}. The narrative modality allows the human supervisor to read about behavior or outcomes in a manner similar to how a human supervisor would interact a human subordinate.

\subsubsection{Values Modality}

The \emph{Values modality} is a numeric presentation of meta-information flow content. This modality can be presented as a percentage (e.g., ``70\%'') \citep{Lai2019OnHP,Wang2016TrustCW}, or as a frequency (e.g., ``2 out of 3'') \citep{Zhang2020EffectOC}. The values modality provides a straightforward mechanism to communicate meta-information amounts. As \citet{Miller2019ExplanationIA} states, however, solely using values may be ``unsatisfying unless accompanied by an underlying causal explanation.''

\subsubsection{Graphical Modality}

The \emph{Graphical modality} is a visual presentation of meta-information flow content. This modality is typically presented as highlights on top of existing content (e.g., highlights of important objects in the field of view), \citep{Wintersberger2021EvaluatingFR}, plots \citep{Bhatt2021UncertaintyAA,McGuirl2006SupportingTC} (e.g., cone-of-uncertainty of a predicted path) \citep{Bhatt2021UncertaintyAA}, or charts \citep{Schaffer2019ICD,Wang2021AreEH} (e.g., bar chart of long-term reward of various outcomes \citep{Schaffer2019ICD}). The graphical modality provides a quickly digestible form of meta-information.

\subsection{Meta-Information Interactivity}

When describing meta-information flows, we need to consider whether interactivity exists or not within the meta-information flow.

Interactivity enables a back-and-forth interaction with the human supervisor to reveal aspects and trends in meta-information flow content. The interactive modality allows the human supervisor to explore the agent’s behavior through adjusting inputs or various internal parameters of the agent. For example, \citet{Cheng2019ExplainingDA} proposed an interactive interface where the human can modify inputs to determine which inputs are driving the agent's decision making. In another example, \citet{Schaffer2019ICD} enabled users to alter estimates of the external environment (i.e., other player tendencies), to understand the relationship between outcome and input.  

\subsection{Meta-Information Timing}

In order to completely describe meta-information flows, we need to also consider the timing of the flow (i.e. when is the flow initiated). The sections below summarize the multiple sub-categories of meta-information flow timings.

\subsubsection{Proactive Timing}

In \emph{Proactive timing}, the meta-information flow is initiated prior to the agent preforming a task. The benefit of Proactive timing is that by informing the human supervisor what is to occur, the supervisor is better prepared to understand agent behavior instead of reacting to behavior that may not be understood in the supervisor’s mental model without the proactive information.  An example of this is described by \citet{Zhu2020EffectsOP}, where they identify proactive explanations as a method of communicating logic rationale for an agent’s actions prior to the agent performing these actions.

\subsubsection{Concurrent Timing}

In \emph{Concurrent timing}, the meta-information flow occurs at the same time as the agent provides the recommendation or course of action. The benefit of Concurrent timing is that the supervisor is informed at the point of having to react to the agent and therefore best leverages the supervisor’s focus.  The majority of the literature provides meta-information concurrently as a one-time flow but can also be dynamically updating as shown by \citet{McGuirl2006SupportingTC}.

\subsubsection{Post-Hoc Timing}

In \emph{Post-Hoc timing}, the meta-information flow is provided after an agent provides a result or performs a course of action. Post-hoc meta-information is provided in response to a request from the human supervisor to explain a previous behavior (e.g., ``Why did/didn’t you do that?,'' ``Why can’t you do that?'') \citep{Graaf2017HowPE,Fox2017ExplainableP} or in the case of a specific trigger, such as anomalous agent behavior \citep{Gregor1999ExplanationsFI}.

\section{Discussion} 

By performing this survey and developing this taxonomy, we have gained insight into the breadth and direction of research in trust calibration from a human-centered point of view. This survey also provided us a unique view into the difficulty of trust calibration and into future directions for our taxonomy.

\subsection{Difficulty of Trust Calibration}

The reason why trust calibration is difficult is that intelligent agents suffer from a form of the Dunning--Kruger Effect \citep{Dunning2011TheDE}: Intelligent agents are too ignorant to know how ignorant they are. More precisely: performing trust calibration forces the intelligent agent to provide cues to a human supervisor indicating the agent’s own output is wrong even when the agent does not know that it is wrong.

\subsubsection{Meta-Information Recommendations}

So what makes an intelligent agent wrong during operations? Over-fitting, not enough training, training not being an accurate representation of operations, poor assumptions or abstractions of the operational environment, the complexity of the operations environment, and unforeseen operations situations are some examples of issues driving an intelligent agent to under-perform in operations. Since we believe these issues drive an agent away from the as-developed performance (e.g., test set accuracy), we recommend providing meta-information content that helps identify a mismatch between the training set and the current instance, or helps identify incorrect modeling assumptions. These include \textit{Uncertainty Content} (especially for determining out-of-distribution) and \textit{Sensitivity Content} in the context of \textit{Graphical Modalities} (especially for highlighting salient features which drive decisions) in \textit{Proactive and Concurrent Timings} for operations.

\subsubsection{Trust-Based Trust Calibration}

Not all operations are the same. In low-stakes, 'everyday' operations (e.g., purchase recommenders), according to \citet{Bunt2012AreEA}, some human supervisors may be wary of any meta-information that requires time and cognitive effort unless it would enhance the speed or accuracy of the system. In high-stakes operations (e.g., military, security, health care, criminal justice), more meta-information may be required to ensure compliance with legal and ethical frameworks. Even though these types of operations are vastly different, we believe both should adapt to human supervisors' level of trust. For example, \citet{Wintersberger2021EvaluatingFR} suggest that trust calibration information should be provided to distrusting supervisors, while not burdening supervisors with trust values that are already high. Therefore we suggest that supervisor trust level be an input to intelligent agents so that the agent can vary the content, modality, timing, interactivity, and amount of trust calibration meta-information accordingly. This construct allows for flexibility in low-cost and high-cost operations as well as providing for cross over cases when a low-cost user wants to understand more or when a high-cost user wants communications brevity (e.g., time-critical military operations).

\subsection{Future Directions}

Now that meta-information content, modalities, interactivity, and timing are formalized in the taxonomy presented in this paper, future studies can determine, for example: which information flow content achieves the best human-agent partnership performance; whether information flow content or modality is what drives this performance; and which information flow content, modalities, interactivity, timings, and amounts work best with which human supervisor characteristics or groups.

One interesting follow on to formalizing meta-information is to standardize this information. Standardization would involve formally defining each content, modality, interactivity, and timing  sub-category including the information formats within each. Even though this may be a daunting task, the benefits could be substantial. For example, having standardized trust calibration meta-information could improve real-world adoption of intelligent agents, reduce the cost of infrastructure necessary to test and field agents, reduce supervisor training, and increase understanding during operations.

Another follow on area deals with meta-information timing. The majority of literature we surveyed on trust calibration and meta-information focused on information flow before, during, or after performing a task, as we described in Section 3.4. Achieving complementary performance, however, may require additional timing phases outside of the operational task. For example, military supervisor-subordinate partnerships develop a ‘shared confidence’ in team members' abilities over time during training exercises where each side of the partnership gains understanding of how the other will react and respond in different situations \citep{ADP6MC}. Initial studies have shown that human-agent partnerships also function better during operations due to training exercises \citep{Johnson2021TheIO} and pre-operations team building\citep{Walliser2019TeamSA}. Therefore, an important area of research would be to determine: what meta-information is needed to support these pre-operations phases and how does the training exercise and rehearsal meta-information differ from what is outlined in this paper.

\section{Limitations}

We chose to limit our survey to trust calibration between a single human and a single agent.  Due to this limitation, trust calibration between humans and multi-agent systems \citep{Setter2017TrustIM} and trust calibration between multiple agents \citep{Liu2016MachineTM} are not discussed.

We also limited our survey to trust calibration meta information in support of individual decisions. Any additional meta-information necessary to support longitudinal trust calibration \citep{Visser2020TowardsAT}, such as reliability \citep{Lyons2021TrustingAS}, was not covered.

We structured our survey based on information flowing from agent to human but acknowledge that there are other very important information flows present in trust calibration that were not covered since they do not originate from the agent. This includes information flows about task risk \citep{Devitt2018TrustworthinessOA}, task complexity \citep{Hancock2011AMO}, and task consequences \citep{Ashoori2019InAW} which change the threshold for trust. 

Finally, since this paper focuses on information, other factors such as agents' physical presence and supervisors' subjective feelings toward intelligent agents \citep{Khavas2021ARO,Hancock2011AMO} are not discussed.

\section{Conclusions}

We have presented a taxonomy based on a review of trust calibration literature for single human to single agent interactions. Through this survey we have distilled the content, modality, interactivity, and timing of meta-information necessary to inform a trust calibration decision for intelligent agents. By focusing on information flows, we have been able to consider several directions to guide future research.

\bibliographystyle{named}
\bibliography{main}

\begin{thebibliography}{}

\bibitem[\protect\citeauthoryear{Arrieta \bgroup \em et al.\egroup
  }{2020}]{Arrieta2020ExplainableAI}
Alejandro~Barredo Arrieta, Natalia D'iaz-Rodr'iguez, Javier~Del Ser, Adrien
  Bennetot, Siham Tabik, A.~Barbado, Salvador Garc'ia, Sergio Gil-L'opez,
  Daniel Molina, Richard Benjamins, Raja Chatila, and Francisco Herrera.
\newblock Explainable artificial intelligence (xai): Concepts, taxonomies,
  opportunities and challenges toward responsible ai.
\newblock {\em ArXiv}, abs/1910.10045, 2020.

\bibitem[\protect\citeauthoryear{Ashoori and Weisz}{2019}]{Ashoori2019InAW}
Maryam Ashoori and Justin~D. Weisz.
\newblock In ai we trust? factors that influence trustworthiness of ai-infused
  decision-making processes.
\newblock {\em ArXiv}, abs/1912.02675, 2019.

\bibitem[\protect\citeauthoryear{Awad \bgroup \em et al.\egroup
  }{2018}]{Awad2018BlamingHI}
Edmond Awad, Sydney Levine, Max Kleiman-Weiner, Sohan Dsouza, Joshua~B.
  Tenenbaum, Azim~F. Shariff, Jean‐François Bonnefon, and Iyad Rahwan.
\newblock Blaming humans in autonomous vehicle accidents: Shared responsibility
  across levels of automation.
\newblock {\em ArXiv}, abs/1803.07170, 2018.

\bibitem[\protect\citeauthoryear{Bhatt \bgroup \em et al.\egroup
  }{2021}]{Bhatt2021UncertaintyAA}
Umang Bhatt, Yunfeng Zhang, Javier Antor{\'a}n, Qingzi~Vera Liao, Prasanna
  Sattigeri, Riccardo Fogliato, Gabrielle~Gauthier Melançon, Ranganath
  Krishnan, Jason Stanley, Omesh Tickoo, Lama Nachman, Rumi Chunara, Adrian
  Weller, and Alice Xiang.
\newblock Uncertainty as a form of transparency: Measuring, communicating, and
  using uncertainty.
\newblock {\em Proceedings of the 2021 AAAI/ACM Conference on AI, Ethics, and
  Society}, 2021.

\bibitem[\protect\citeauthoryear{Bode and Watts}{2021}]{Bode2021AlreadyWrong}
Ingvild Bode and Tom Watts.
\newblock Worried about the autonomous weapons of the future? look at what's
  already gone wrong.
\newblock
  \url{https://thebulletin.org/2021/04/worried-about-the-autonomous-weapons-of-the-future-\\look-at-whats-already-gone-wrong/},
  2021.
\newblock Last Accessed: 2022-02-14.

\bibitem[\protect\citeauthoryear{Bunt \bgroup \em et al.\egroup
  }{2012}]{Bunt2012AreEA}
Andrea Bunt, Matthew Lount, and Catherine Lauzon.
\newblock Are explanations always important?: a study of deployed, low-cost
  intelligent interactive systems.
\newblock In {\em IUI '12}, 2012.

\bibitem[\protect\citeauthoryear{Bussone \bgroup \em et al.\egroup
  }{2015}]{Bussone2015TheRO}
Adrian Bussone, Simone Stumpf, and Dympna~M. O'Sullivan.
\newblock The role of explanations on trust and reliance in clinical decision
  support systems.
\newblock {\em 2015 International Conference on Healthcare Informatics}, pages
  160--169, 2015.

\bibitem[\protect\citeauthoryear{Cai \bgroup \em et al.\egroup
  }{2019}]{Cai2019TheEO}
Carrie~J. Cai, Jonas Jongejan, and Jess Holbrook.
\newblock The effects of example-based explanations in a machine learning
  interface.
\newblock {\em Proceedings of the 24th International Conference on Intelligent
  User Interfaces}, 2019.

\bibitem[\protect\citeauthoryear{Cheng \bgroup \em et al.\egroup
  }{2019}]{Cheng2019ExplainingDA}
Hao~Fei Cheng, Ruotong Wang, Zheng Zhang, Fiona O'Connell, Terrance Gray,
  F.~Maxwell Harper, and Haiyi Zhu.
\newblock Explaining decision-making algorithms through ui: Strategies to help
  non-expert stakeholders.
\newblock {\em Proceedings of the 2019 CHI Conference on Human Factors in
  Computing Systems}, 2019.

\bibitem[\protect\citeauthoryear{de Graaf and Malle}{2017}]{Graaf2017HowPE}
Maartje de~Graaf and Bertram~F. Malle.
\newblock How people explain action (and autonomous intelligent systems should
  too).
\newblock In {\em AAAI Fall Symposia}, 2017.

\bibitem[\protect\citeauthoryear{de Visser \bgroup \em et al.\egroup
  }{2020}]{Visser2020TowardsAT}
Ewart de~Visser, Marieke M.~M. Peeters, Malte~F. Jung, Spencer Kohn, Tyler~H.
  Shaw, Richard Pak, and Mark~A. Neerincx.
\newblock Towards a theory of longitudinal trust calibration in human–robot
  teams.
\newblock {\em International Journal of Social Robotics}, 12:459--478, 2020.

\bibitem[\protect\citeauthoryear{Devitt}{2018}]{Devitt2018TrustworthinessOA}
Susannah~Kate Devitt.
\newblock Trustworthiness of autonomous systems.
\newblock In {\em CDC 2018}, 2018.

\bibitem[\protect\citeauthoryear{Dietvorst \bgroup \em et al.\egroup
  }{2015}]{Dietvorst2015AlgorithmAP}
Berkeley~J. Dietvorst, Joseph~P. Simmons, and Cade Massey.
\newblock Algorithm aversion: people erroneously avoid algorithms after seeing
  them err.
\newblock {\em Journal of experimental psychology. General}, 144 1:114--26,
  2015.

\bibitem[\protect\citeauthoryear{Dodge \bgroup \em et al.\egroup
  }{2019}]{Dodge2019ExplainingMA}
Jonathan Dodge, Qingzi~Vera Liao, Yunfeng Zhang, Rachel K.~E. Bellamy, and
  Casey Dugan.
\newblock Explaining models: an empirical study of how explanations impact
  fairness judgment.
\newblock {\em Proceedings of the 24th International Conference on Intelligent
  User Interfaces}, 2019.

\bibitem[\protect\citeauthoryear{Dunning}{2011}]{Dunning2011TheDE}
David Dunning.
\newblock The dunning–kruger effect: On being ignorant of one’s own
  ignorance.
\newblock In {\em Advances in Experimental Social Psychology, Volume 44,
  Academic Press}, 2011.

\bibitem[\protect\citeauthoryear{Ehrlich \bgroup \em et al.\egroup
  }{2011}]{Ehrlich2011TakingAF}
Kate Ehrlich, Susanna~E. Kirk, John~F. Patterson, Jamie~C. Rasmussen, Steven~I.
  Ross, and Dan Gruen.
\newblock Taking advice from intelligent systems: the double-edged sword of
  explanations.
\newblock In {\em IUI '11}, 2011.

\bibitem[\protect\citeauthoryear{Ehsan \bgroup \em et al.\egroup
  }{2021a}]{Ehsan2021ExpandingET}
Upol Ehsan, Qingzi~Vera Liao, Michael~J. Muller, Mark~O. Riedl, and Justin~D.
  Weisz.
\newblock Expanding explainability: Towards social transparency in ai systems.
\newblock {\em Proceedings of the 2021 CHI Conference on Human Factors in
  Computing Systems}, 2021.

\bibitem[\protect\citeauthoryear{Ehsan \bgroup \em et al.\egroup
  }{2021b}]{Ehsan2021TheWI}
Upol Ehsan, Samir Passi, Qingzi~Vera Liao, Larry Chan, I-Hsiang Lee, Michael~J.
  Muller, and Mark~O. Riedl.
\newblock The who in explainable ai: How ai background shapes perceptions of ai
  explanations.
\newblock {\em ArXiv}, abs/2107.13509, 2021.

\bibitem[\protect\citeauthoryear{Fox \bgroup \em et al.\egroup
  }{2017}]{Fox2017ExplainableP}
Maria Fox, Derek Long, and Daniele Magazzeni.
\newblock Explainable planning.
\newblock {\em ArXiv}, abs/1709.10256, 2017.

\bibitem[\protect\citeauthoryear{Gregor and
  Benbasat}{1999}]{Gregor1999ExplanationsFI}
Shirley~D Gregor and Izak Benbasat.
\newblock Explanations from intelligent systems: Theoretical foundations and
  implications for practice.
\newblock {\em MIS Q.}, 23:497--530, 1999.

\bibitem[\protect\citeauthoryear{Guidotti \bgroup \em et al.\egroup
  }{2019}]{Guidotti2019ASO}
Riccardo Guidotti, Anna Monreale, Franco Turini, Dino Pedreschi, and Fosca
  Giannotti.
\newblock A survey of methods for explaining black box models.
\newblock {\em ACM Computing Surveys (CSUR)}, 51:1 -- 42, 2019.

\bibitem[\protect\citeauthoryear{Hancock \bgroup \em et al.\egroup
  }{2011}]{Hancock2011AMO}
Peter~A. Hancock, Deborah~R. Billings, Kristin~E. Schaefer, Jessie~Y.C. Chen,
  Ewart de~Visser, and Raja Parasuraman.
\newblock A meta-analysis of factors affecting trust in human-robot
  interaction.
\newblock {\em Human Factors: The Journal of Human Factors and Ergonomics
  Society}, 53:517 -- 527, 2011.

\bibitem[\protect\citeauthoryear{Huysmans \bgroup \em et al.\egroup
  }{2011}]{Huysmans2011AnEE}
Johan Huysmans, Karel Dejaeger, Christophe Mues, Jan Vanthienen, and Bart
  Baesens.
\newblock An empirical evaluation of the comprehensibility of decision table,
  tree and rule based predictive models.
\newblock {\em Decis. Support Syst.}, 51:141--154, 2011.

\bibitem[\protect\citeauthoryear{Johnson \bgroup \em et al.\egroup
  }{2021}]{Johnson2021TheIO}
Craig~J. Johnson, Mustafa Demir, Nathan~J. Mcneese, Jamie~C. Gorman,
  Alexandra~T. Wolff, and Nancy~J. Cooke.
\newblock The impact of training on human-autonomy team communications and
  trust calibration.
\newblock {\em Human factors}, page 187208211047323, 2021.

\bibitem[\protect\citeauthoryear{Khavas}{2021}]{Khavas2021ARO}
Zahra~Rezaei Khavas.
\newblock A review on trust in human-robot interaction.
\newblock {\em ArXiv}, abs/2105.10045, 2021.

\bibitem[\protect\citeauthoryear{Kunkel \bgroup \em et al.\egroup
  }{2019}]{Kunkel2019LetME}
Johannes Kunkel, Tim Donkers, Lisa Michael, Catalin-Mihai Barbu, and J{\"u}rgen
  Ziegler.
\newblock Let me explain: Impact of personal and impersonal explanations on
  trust in recommender systems.
\newblock {\em Proceedings of the 2019 CHI Conference on Human Factors in
  Computing Systems}, 2019.

\bibitem[\protect\citeauthoryear{Lai and Tan}{2019}]{Lai2019OnHP}
Vivian Lai and Chenhao Tan.
\newblock On human predictions with explanations and predictions of machine
  learning models: A case study on deception detection.
\newblock {\em Proceedings of the Conference on Fairness, Accountability, and
  Transparency}, 2019.

\bibitem[\protect\citeauthoryear{Lee and See}{2004}]{Lee2004TrustIA}
John~D. Lee and Katrina~A. See.
\newblock Trust in automation: Designing for appropriate reliance.
\newblock {\em Human Factors: The Journal of Human Factors and Ergonomics
  Society}, 46:50 -- 80, 2004.

\bibitem[\protect\citeauthoryear{Lee}{2018}]{Lee2018Hulu}
Timothy~B. Lee.
\newblock Police: Uber driver was streaming hulu just before fatal self-driving
  car crash.
\newblock
  \url{https://arstechnica.com/cars/2018/06/police-uber-driver-was-streaming-hulu-just-before-fatal\\self-driving-car-crash/},
  2018.
\newblock Last Accessed: 2022-02-07.

\bibitem[\protect\citeauthoryear{Liu \bgroup \em et al.\egroup
  }{2016}]{Liu2016MachineTM}
Ling Liu, Margaret~L. Loper, Yusuf {\"O}zkaya, Abdurrahman Yaşar, and Emre
  Yigitoglu.
\newblock Machine to machine trust in the iot era.
\newblock In {\em TRUST@AAMAS}, 2016.

\bibitem[\protect\citeauthoryear{Lowens}{2020}]{Lowens2020AlgAver}
Ethan Lowens.
\newblock Accuracy is not enough: The task mismatch explanation of algorithm
  aversion and its policy implications.
\newblock {\em Harvard Journal of Law \& Technology, Vol. 34, No. 1}, 2020.

\bibitem[\protect\citeauthoryear{Lyell and Coiera}{2017}]{Lyell2017Automation}
David Lyell and Enrico~W. Coiera.
\newblock Automation bias and verification complexity: a systematic review.
\newblock {\em Journal of the American Medical Informatics Association},
  24:423–431, 2017.

\bibitem[\protect\citeauthoryear{Lyons \bgroup \em et al.\egroup
  }{2021}]{Lyons2021TrustingAS}
Joseph~B. Lyons, Thy Vo, Kevin~T. Wynne, Sean Mahoney, Chang~Soo Nam, and Darci
  Gallimore.
\newblock Trusting autonomous security robots: The role of reliability and
  stated social intent.
\newblock {\em Human Factors: The Journal of Human Factors and Ergonomics
  Society}, 63:603 -- 618, 2021.

\bibitem[\protect\citeauthoryear{McGuirl and
  Sarter}{2006}]{McGuirl2006SupportingTC}
John~M. McGuirl and Nadine~B. Sarter.
\newblock Supporting trust calibration and the effective use of decision aids
  by presenting dynamic system confidence information.
\newblock {\em Human Factors: The Journal of Human Factors and Ergonomics
  Society}, 48:656 -- 665, 2006.

\bibitem[\protect\citeauthoryear{Miller}{2019}]{Miller2019ExplanationIA}
Tim Miller.
\newblock Explanation in artificial intelligence: Insights from the social
  sciences.
\newblock {\em Artif. Intell.}, 267:1--38, 2019.

\bibitem[\protect\citeauthoryear{Muir}{1987}]{Muir1987Trust}
Bonnie~M. Muir.
\newblock Trust between humans and machines, and the design of decision aids.
\newblock {\em Int. J. Man Mach. Stud.}, 27:527--539, 1987.

\bibitem[\protect\citeauthoryear{Schaffer \bgroup \em et al.\egroup
  }{2019}]{Schaffer2019ICD}
James~Austin Schaffer, John O'Donovan, James~R. Michaelis, Adrienne~Jeanisha
  Raglin, and Tobias H{\"o}llerer.
\newblock I can do better than your ai: expertise and explanations.
\newblock {\em Proceedings of the 24th International Conference on Intelligent
  User Interfaces}, 2019.

\bibitem[\protect\citeauthoryear{Setter \bgroup \em et al.\egroup
  }{2017}]{Setter2017TrustIM}
Tina Setter, Andrea Gasparri, and Magnus Egerstedt.
\newblock Trust in multi-agent networks: From self-centered to team-oriented.
\newblock {\em 2017 American Control Conference (ACC)}, pages 997--1002, 2017.

\bibitem[\protect\citeauthoryear{Sperrle \bgroup \em et al.\egroup
  }{2020}]{Sperrle2020ShouldWT}
Fabian Sperrle, Mennatallah El-Assady, Grace Guo, Duen~Horng Chau, Alex Endert,
  and Daniel~A. Keim.
\newblock Should we trust (x)ai? design dimensions for structured experimental
  evaluations.
\newblock {\em ArXiv}, abs/2009.06433, 2020.

\bibitem[\protect\citeauthoryear{Svrcek \bgroup \em et al.\egroup
  }{2019}]{Svrcek2019TowardsUP}
Martin Svrcek, Michal Kompan, and M{\'a}ria Bielikov{\'a}.
\newblock Towards understandable personalized recommendations: Hybrid
  explanations.
\newblock {\em Comput. Sci. Inf. Syst.}, 16:179--203, 2019.

\bibitem[\protect\citeauthoryear{Tomsett \bgroup \em et al.\egroup
  }{2020}]{Tomsett2020RapidTC}
Richard~J. Tomsett, Alun~David Preece, Dave Braines, Federico Cerutti, Supriyo
  Chakraborty, Mani~B. Srivastava, Gavin Pearson, and Lance~M. Kaplan.
\newblock Rapid trust calibration through interpretable and uncertainty-aware
  ai.
\newblock {\em Patterns}, 1, 2020.

\bibitem[\protect\citeauthoryear{USArmy}{2019}]{ADP6MC}
USArmy.
\newblock Army doctrine publication (adp) 6-0, mission command.
\newblock
  \url{https://armypubs.army.mil/ProductMaps/PubForm/Details.aspx?PUB_ID=1007502},
  2019.
\newblock Last Accessed: 2022-02-09.

\bibitem[\protect\citeauthoryear{Walliser \bgroup \em et al.\egroup
  }{2019}]{Walliser2019TeamSA}
James Walliser, Ewart~J. de~Visser, Eva Wiese, and Tyler~H. Shaw.
\newblock Team structure and team building improve human–machine teaming with
  autonomous agents.
\newblock {\em Journal of Cognitive Engineering and Decision Making}, 13:258 --
  278, 2019.

\bibitem[\protect\citeauthoryear{Wang and Yin}{2021}]{Wang2021AreEH}
Xinru Wang and Ming Yin.
\newblock Are explanations helpful? a comparative study of the effects of
  explanations in ai-assisted decision-making.
\newblock {\em 26th International Conference on Intelligent User Interfaces},
  2021.

\bibitem[\protect\citeauthoryear{Wang \bgroup \em et al.\egroup
  }{2016}]{Wang2016TrustCW}
Ning Wang, David~V. Pynadath, and Susan~G. Hill.
\newblock Trust calibration within a human-robot team: Comparing automatically
  generated explanations.
\newblock {\em 2016 11th ACM/IEEE International Conference on Human-Robot
  Interaction (HRI)}, pages 109--116, 2016.

\bibitem[\protect\citeauthoryear{Wintersberger \bgroup \em et al.\egroup
  }{2021}]{Wintersberger2021EvaluatingFR}
Philipp Wintersberger, Frederica Janotta, Jakob Peintner, Andreas L{\"o}cken,
  and Andreas Riener.
\newblock Evaluating feedback requirements for trust calibration in automated
  vehicles.
\newblock {\em it - Information Technology}, 63:111 -- 122, 2021.

\bibitem[\protect\citeauthoryear{Yang \bgroup \em et al.\egroup
  }{2021}]{Yang2021GeneralizedOD}
Jingkang Yang, Kaiyang Zhou, Yixuan Li, and Ziwei Liu.
\newblock Generalized out-of-distribution detection: A survey.
\newblock {\em ArXiv}, abs/2110.11334, 2021.

\bibitem[\protect\citeauthoryear{Zhang \bgroup \em et al.\egroup
  }{2020}]{Zhang2020EffectOC}
Yunfeng Zhang, Qingzi~Vera Liao, and Rachel K.~E. Bellamy.
\newblock Effect of confidence and explanation on accuracy and trust
  calibration in ai-assisted decision making.
\newblock {\em Proceedings of the 2020 Conference on Fairness, Accountability,
  and Transparency}, 2020.

\bibitem[\protect\citeauthoryear{Zhu and Williams}{2020}]{Zhu2020EffectsOP}
Lixiao Zhu and Thomas~Emrys Williams.
\newblock Effects of proactive explanations by robots on human-robot trust.
\newblock In {\em ICSR}, 2020.

\end{thebibliography}

\end{document}